\documentclass[]{elsart5p}
\usepackage{graphicx}
\usepackage[fleqn]{amsmath}
\usepackage{subfigure}
\newcommand{\uvec}{\mathbf{u}}

\newcommand{\mom}{$\langle \rho u_z\rangle\:$}
\newcommand{\modmom}{$\langle|\rho u_z|\rangle\:$}
\newcommand{\bsq}{$\langle B^2\rangle\:$}

\begin{document}
\begin{frontmatter}
\title{Three-layer magnetoconvection}
\author[a,b]{M.-K.\ Lin\ead{mkl23@cam.ac.uk}},
\author[a]{L.\ J.\ Silvers\ead{ljs53@damtp.cam.ac.uk}\corauthref{cor}},
\author[a]{M.\ R.\ E.\ Proctor\ead{mrep@damtp.cam.ac.uk}}
\corauth[cor]{Corresponding author.}
\address[a]{Department of Applied Mathematics and Theoretical Physics,
  University of Cambridge, Cambridge, CB3 OWA, United Kingdom}
\address[b]{St.\ Catharine's College, University of Cambridge, Cambridge, CB2 1RL, United Kingdom}

\begin{abstract}
It is believed that some stars have two or more convection zones in
close proximity near to the stellar photosphere. These zones are
separated by convectively stable regions that are relatively narrow.
Due to the close proximity of these regions it is important to
construct mathematical models to understand the transport and mixing
of passive and dynamic quantities. One key quantity of interest is a
magnetic field, a dynamic vector quantity, that can drastically
alter the convectively driven flows, and have an important role in coupling the different layers. In this paper we present the
first investigation into the effect of an imposed magnetic field in such a geometry. We focus our attention on the effect of field strength and
show that, while there are some similarities with results for magnetic
field evolution in a single layer, new and interesting phenomena are also
present in a three layer system.
\end{abstract}
\end{frontmatter}
\section{Introduction}\label{intro}
\indent Throughout the Universe there are a plethora of stars with a
variety of different internal structures\cite{schwarzschild65}.
Amongst the stars that we observe there are some, such as A-type stars,
which are believed to have  multiple convection zones near the
surface
 \cite{landstreet98, silaj05}, which is a phenomenon 
that results, to some extent, from the non-trivial changes in the chemical makeup as a function of distance centre
of the star is increased \cite{toomre76}. The convection
zones in these stars are thin, as compared to the radius of the star
but are important as they affect the transport properties of this
part of the star.

As with all cases of convection in an astrophysical context, there
are no solid boundaries encasing the convectively unstable fluid. Thus the ascending and descending plumes in the unstable regions can overshoot the
convectively unstable layer and continue into the adjacent
convectively stable region. Indeed, if the convection is
sufficiently strong, or the adjacent stable region is sufficiently
narrow, the overshooting plumes can pass straight through the stable
region and enter the second convectively unstable region. It is thus clear that 
fascinating dynamical behaviour can be envisioned for this system and it is
important to study such systems if we are to understand transport
and mixing in stars where more than one convection zone is present.

Early analytical work on convection in stars with multiple convection zones indicated that
a separation of more than two pressure scale heights between the convection allows them to be considered as disjoint
\cite{toomre76,latour76,latour81}. With advances in computational
resources, it has since become possible carry out direct simulations
of convection zones and their interaction with radiative zones, in
application to solar convection or multiple convection zones in
A-stars \cite{MWFL,silvers07}. These simulations
show the importance of further investigations into the mixing
and transport in these stars as they demonstrate that a large
degree of separation is required for the convection zones to be
considered dynamically and thermally isolated\cite{silvers07}.

The numerical investigations to date have been aimed at providing a
solid basis for later, more complex, models. There are many further
aspects of the physics of these stars which need to be considered and questions that still remain.
Amongst these is the fact that convectively unstable regions in such
stars are permeated by a magnetic field \cite{preston74}.

There has, to date, been an extensive literature concerning the effect of
a magnetic field on a convectively unstable layer (see, for example,
\cite{weiss66,hurlburt88,matthews95,proctor05}) or in a convectively unstable
layer that abuts onto a single convectively stable layer (see, for
example, \cite{tobias:s}). However as yet there has been no examination of the evolution of a magnetic field in a scenario with multiple
convection layers as described by Silvers \& Proctor
\cite{silvers07}. The purely hydrodynamic problem proved not to be a simple extension of
single-layer systems, and we naturally anticipate at least the same complexity 
once a magnetic field is included. Exploring the effect of a
magnetic field is also of interest because it has been conjectured
that certain chemical anomalies could result from magnetic fields in
stellar atmospheres \cite{preston74,vauclair82}. Michaud
\cite{michaud70} suggested that field lines might stabilize the atmosphere
to allow diffusion and guide particles into patches. It has also
been suggested that magnetic fields may reduce the ion diffusion
velocity \cite{vauclair82}.

In magnetoconvection calculations in a single unstable layer, the state that is reached after long times depends strongly on the strength of the magnetic field permeating the system. We expect to see a similar sensitivity here, and in addition we expect that the coupling between the layers is strongly affected by the field. Thus in the present paper we will
explore the effect of varying field strength on the convection and
interaction between two layers.

In this work we consider an atmosphere with two convective zones separated by a stable
layer with an initially vertical magnetic field.  We do not address
the specific problem of chemical anomalies by detailed modelling of
stellar atmosphere composition and diffusion, as our goal is to
provide a first understanding the effect of varying the strength of
the magnetic field on convection through a simple model.

This paper is organised as follows: in the next section we describe
our model with relevant equations, parameters and numerical method.
In section \ref{results} we present the results for cases with
different strength magnetic fields. Finally, in section \ref{conc}
we summarize our findings.

\section{Model}

We consider an atmosphere taking the form of a compressible fluid in a slab, with temperature decreasing piecewise linearly with height, permeated
by an imposed vertical magnetic field. The slab is comprised of three layers of equal
thickness, the top and bottom being convectively unstable and the
middle stable.

Apart from the multi-layer feature of the geometry, the equations are in standard form, as described in \cite{silvers07,matthews95}. The governing equations are given in dimensionless form; lengths are scaled by the depth $d$ of each layer; density and temperature by $\rho_0$
and $T_0$, (values at $z=0$, where $z$ increases downwards);
times by the sound crossing time $d/\sqrt{R_*T_0}$ where $R_*$ is the
gas constant; and magnetic field by $B_0$, the magnitude of the initial
uniform field. The equations then take the form:
\begin{equation}
\frac{\partial\rho}{\partial t}+\nabla\cdot(\rho\uvec)=0
\end{equation}
\begin{eqnarray}
\rho\left(\frac{\partial\uvec}{\partial t} +
\uvec\cdot\nabla\uvec\right) =
&-&\nabla(P+FB^2/2)+\theta(m+1)\rho\hat{\bf{z}} \nonumber\\ &+&
\nabla\cdot(F\mathbf{B}\mathbf{B}+\rho\sigma\kappa\mbox{\boldmath$\tau$})
\end{eqnarray}
\begin{eqnarray}
\frac{\partial T}{\partial t}+\mathbf{u}\cdot\nabla
T&+&(\gamma-1)T\nabla\cdot\uvec=\frac{\gamma\kappa}{\rho}\nabla^2T
\nonumber \\ &+& \kappa(\gamma-1)(\sigma\tau^2/2+F\zeta_0J^2/\rho)
\end{eqnarray}
\begin{equation}
\frac{\partial\bf{B}}{\partial
t}=\nabla\wedge(\uvec\wedge\mathbf{B}-\zeta_0\kappa\nabla\wedge\bf{B})
\end{equation}
\begin{equation}
\nabla\cdot\mathbf{B}=0
\end{equation}
\begin{equation}
 P=\rho T
\end{equation}
here $F=B_0^2/(R_*T_0\rho_0\mu_0)$, $\kappa=K/(d\rho_0
c_p\sqrt{R_*T_0})$ the dimensionless thermal diffusivity,
$\tau_{ij}\equiv\partial_ju_i+\partial_iu_j-(2/3)\delta_{ij}\partial_ku_k$
is the stress tensor and $\zeta_0=\eta c_p\rho_0/K$ where $\eta$ is
the magnetic diffusivity. Other
quantities have their usual meanings. The equations are solved using
a mixed finite-difference/pseudospectral code. More details on the
numerical method and code may be found in \cite{matthews95}.
Throughout this paper we will use a resolution of
$240\times64\times64$.

For convenience we define the Chandrasekhar number
$Q={F}/{\zeta_0\sigma\kappa^2}$,
which provides a measure of field strength relative to diffusion and
in what follows we will focus on the effect of varying this quantity, with other parameters held fixed. Their values are
given in Table \ref{param}. Note that, for simplicity, we will
introduce the notation that subscripts 1, 2 and 3  refer to
respectively the top, middle and bottom zones. Also, we note here
that our choice of polytropic indices corresponds to the stiffness
parameters $S_1=S_3=-1.0$ for the top and bottom and $S_2=5.0$ for
the middle layer, where $S_2=(m_2-m_{ad})/(m_{ad}-m_1)$ and $
S_3=(m_3-m_{ad})(m_{ad}-m_1)$; see e.g. \cite{silvers07}.
\begin{table}[!h]
\caption{Parameter values.}\label{param}
\begin{center}
\begin{tabular*}{0.48\textwidth}{@{\extracolsep{\fill}}lll}
\hline Symbol & Name & Value\\ \hline $z_{m}$ &Vertical
extent&3.0\\ $y_{m}=x_{m}$&Horizontal extent&8.0\\ $\gamma$&Ratio
of specific heats&5/3\\ $\sigma$&Prandtl number ($=\mu c_p/K$, viscosity $\mu$)&1.0\\ $\theta$&Temperature
difference across a layer&10\\ $\zeta_0$&Magnetic
diffusivity&0.2\\ $m_1=m_3$& Top and bottom polytropic index&1.0\\
$m_2$&Middle polytropic index&4.0\\ $R_1$&Rayleigh number near the
top&5000.0\\ $Q$&Chandrasekhar number&variable\\
\hline
\end{tabular*}
\end{center}
\end{table}

The initial three-layer structure, with different polytropic indices in the three layers is obtained by choosing a thermal
conductivity profile of the form \cite{silvers07}:
\begin{eqnarray}\label{thermcond}
K & = & \frac{K_1}{2} \Big[1+\frac{K_2+K_3}{K_1}-\tanh
\Big(\frac{z-1}{\Delta}\Big) \nonumber\\ &+ &
\frac{K_3}{K_1}\tanh\Big(\frac{z-2}{\Delta}\Big) \nonumber \\
&-&\frac{K_2}{K_1}\tanh\big(\frac{z-2}{\Delta}\Big)\tanh\big(\frac{z-1}{\Delta}\Big)\Big]
\end{eqnarray}
where $\Delta=0.1$ in this case, so as to allow a smooth transition between the layers. To the static state we add random
velocity perturbations in the range $[-0.05,0.05]$ and allow the
system to evolve. The boundary conditions at the top and bottom of
the domain are taken to be:
\begin{eqnarray}\label{BC}
T=1,\:u_z=\frac{\partial u_x}{\partial z}=B_x=B_y=\frac{\partial B_z}{\partial z}=0 \:\text{at}\: z=0, \nonumber\\
\frac{\partial T}{\partial z}=\theta,\: u_z =\frac{\partial
u_x}{\partial z}=B_x=B_y=\frac{\partial B_z}{\partial
z}=0\:\text{at}\: z=3,
\end{eqnarray}
and all quantities are taken to be periodic in $x$ and $y$ with periods $x_m,y_m$.

\section{Results}\label{results}

\begin{figure}[!h]
\includegraphics[width=0.9\linewidth,clip,trim=0.25cm 5cm 0.5cm 1cm]{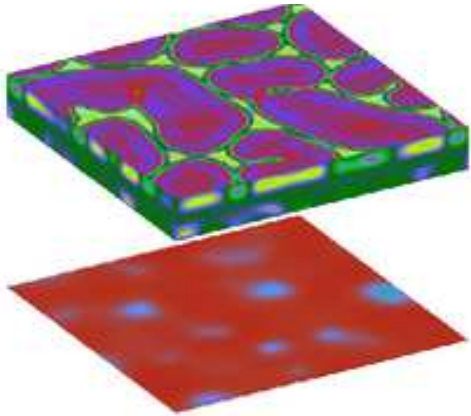}
\caption{Relative distribution of vertical component of magnetic field ( near the top and
  bottom) and vertical component of momentum (sides), for the case $Q=100$
  at $t=29.58$.}\label{mhd2_3d}
\end{figure}

In this paper we explore the effect of varying magnetic field
strength, by varying the Chandrasekhar number, $Q$. We begin with a  discussion of the  weak field case where $Q=100$. Figure
\ref{mhd2_3d} shows the distributions of vertical momentum density
($\rho u_z$, sides of the box) and of vertical component of magnetic
field ($B_z$ near the top and bottom) once the motion is fully
established. This figure shows that the vertical magnetic field structure is dominated by regions of width between 0.4-0.7 between the convection cells in the upper layer. The lower layer does not resemble the upper layer, in spite of having the same polytropic index, because it has greater density and different values of other physical properties.

The bottom field is relatively weak and much more uniform, the most prominent
structures being rising convergent plumes (diameter $\simeq0.9$) with slightly enhanced values of 
$B_z$. Distinct upflow and downflow regions can be seen in the upper layer. In the central, stably stratified layer
where $|\rho u_z|$ is small, 
$B_z$is almost uniform . The lower convection zone, in contrast to the upper layer, has fewer and less
ordered convection cells, and there is little correlation with  the field in the upper convection layer. 
\begin{figure}
\begin{center}
\includegraphics[width=.5\linewidth,clip,trim=0.5cm 6.5cm 0cm 0.75cm]{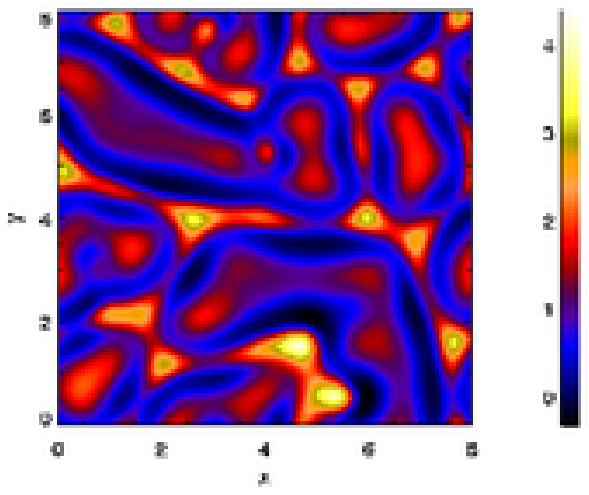}\includegraphics[width=.5\linewidth,clip,trim=0.5cm 6.5cm 0cm 0.75cm]{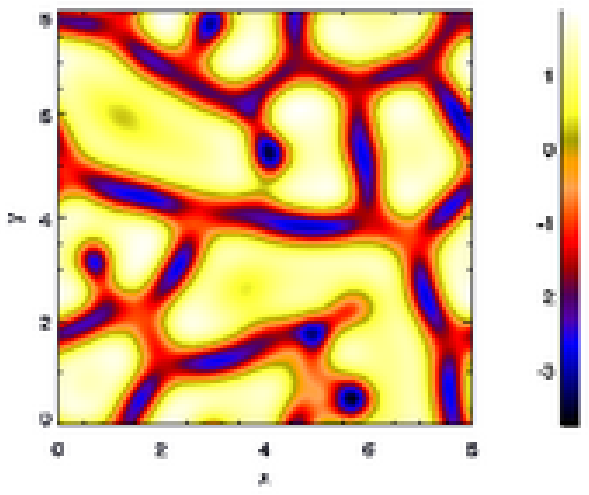}\\
\includegraphics[width=.5\linewidth,clip,trim=0.5cm 6.5cm 0cm 0.75cm]{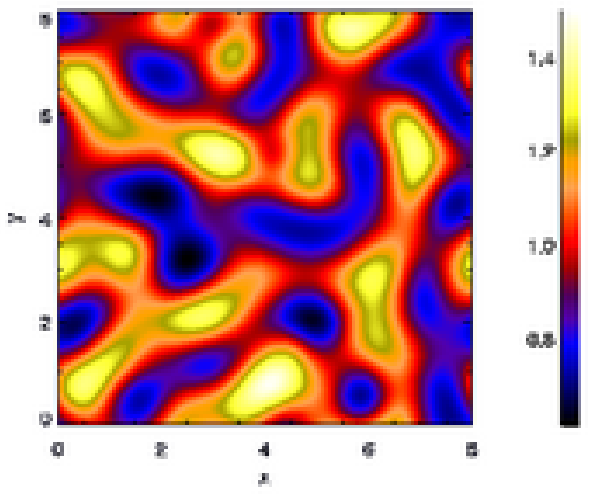}\includegraphics[width=.5\linewidth,clip,trim=0.5cm 6.5cm 0cm 0.75cm]{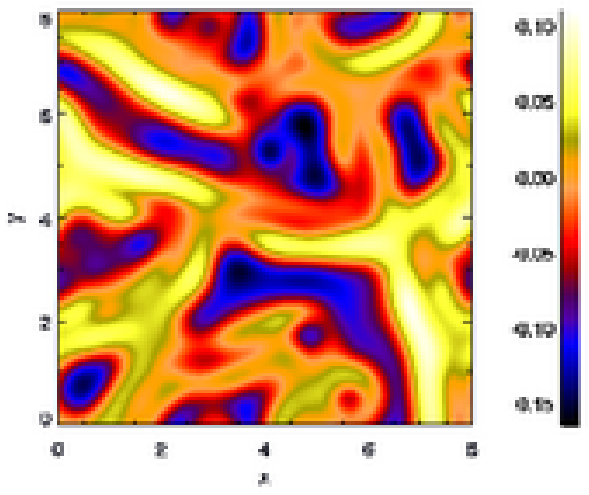}\\
\includegraphics[width=.5\linewidth,clip,trim=0.5cm 6.5cm 0cm 0.75cm]{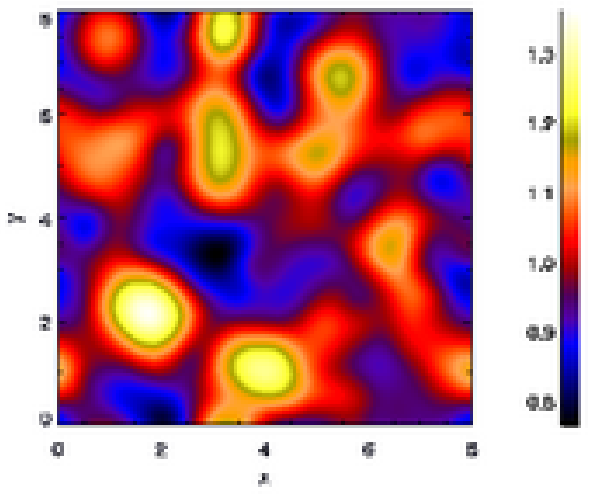}\includegraphics[width=.5\linewidth,clip,trim=0.5cm 6.5cm 0cm 0.75cm]{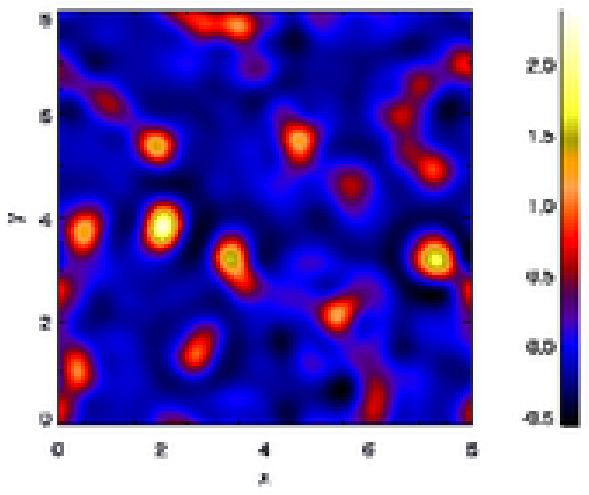}\\
\end{center}
\caption{From top to bottom: vertical component of magnetic field
(left) and vertical component of momentum (right) in the horizontal
plane
  at $z=0.75, 1.5, 2.25$; for $Q=100$ and $t=29.58$. }\label{mhd2_mag_layer}
\end{figure}

To explore the change in flows in more detail we consider Figure
\ref{mhd2_mag_layer} that shows horizontal slices of $B_z$ and $\rho u_z$ at the middle of each of the zones. At $z=0.75$,
regions of high $B_z$ corresponds to vertical motion, and from the
colourbar range on the $\rho u_z$ plot we see that downflows are
stronger, consistent with previous studies of compressible
magnetoconvection \cite{hurlburt88}. Regions of weakest $B_z$
matches to where $|\rho u_z|\sim0$ so any motion is in the
horizontal plane. This is again consistent with previous
investigations, which showed that magnetic flux is swept by convection into converging
regions within which the field is nearly parallel to the fluid motion
\cite{weiss66}. We also note that in the $\rho u_z$ plots, there is
little variation within the upflow cells.

In the convectively stable region, at $z=1.5$, there is a much smaller variation $B_z$ and $\rho u_z$
than in the upper convection zones. The pattern of motion
is very weakly correlated with that at at $z=0.75$ for $B_z$
with rolls still dominant but of larger widths ($\sim 1$ unit). The
mid-layer $\rho u_z$ is typically anti-correlated to the upper
layer; for example the downflow region  in the lower half of the
plot corresponds to upflow  at $z=0.75$, although the former is
thinner in extent. Comparing the plots we can see that vertical field and motion at $z=0.5$ are almost unrelated. It is important to note that no
convection can occur in the middle of the box because of our choice
of polytropic index; so any motion must be due to overshooting
plumes from either convection zone, and it would seem that the magnetic field pattern is due almost entirely to the vigorous convection in the unstable layers.

In the lower
convection zone, at $z=2.25$, the contrast in $B_z$ is similar to that in the central region but the pattern is more cellular. Interestingly, although
this layer is convectively unstable, $B_z$ does not correlate well
to $\rho u_z$, unlike in the top layer. The distribution of $\rho u_z$
is almost uniform with small cells (diameter $\sim 0.5$ units) of
strong upwards motion and their positions appear unrelated. These
slice plots show distinct changes in $B_z$ across the layers,
suggesting that for a weak field, its associated structure can not
be easily communicated across boundaries, from this perspective the
three layers appear independent. However, the
boundary conditions on the interface allow overshooting, which is another form of
communication across boundaries, and is best illustrated by
considering the variation of $|\rho u_z|$ with $z$.

\begin{figure}
\includegraphics[scale=.50,clip,trim=2cm 13cm 0cm 0cm]{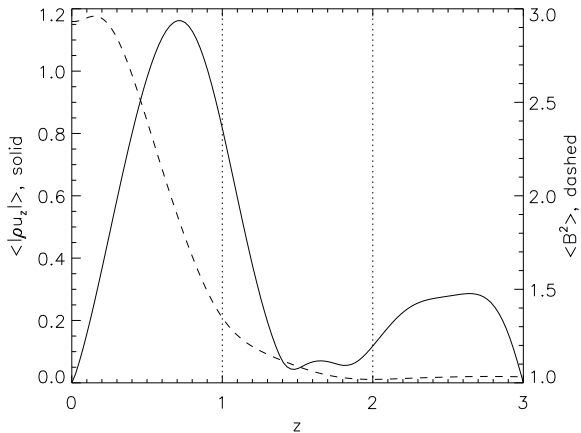}
\caption{Horizontal average of modulus of vertical momentum density (solid) and
   magnetic energy  (solid line), as a function of
  depth, for $Q=100$ at $t=29.58$.}\label{mhd2_layer}
\end{figure}

Fig. \ref{mhd2_layer} shows snapshots of\modmom and \bsq as a function of $z$
for the $Q=100$ case, where angle brackets denote horizontal
averages. It is possible that such snapshots can be misleading as they
can be contaminated by acoustic and gravity modes. However, we  have verified by looking at other snapshots that the distributions of the two quantities shown are typical in the statistically steady state.  As expected vertical motion dominates in the two
convectively unstable zones due to convection, but the solid lines
extend from both unstable zones into the middle so there is non-zero
vertical motion throughout the stable region which indicate
overshooting. The motion in the upper convection zone is more vigorous and the
solid curve extends into the mid-layer more than that from the lower
convection zone, which suggests more overshooting from the upper
layer into the middle. This is indeed consistent with the slice
plots (Figure \ref{mhd2_mag_layer}); but as we will show later, the
correspondence is not universal. In the statistically steady state the top
also contains most of the magnetic energy. The typical value of \bsq
in the middle is $\sim0.37$ times the maximum (in the upper layer)
so some of the perturbation to magnetic energy `overflows' into the
middle. Although there is more motion in the lower zone than the middle
there is not much field amplification, and from this together with the
slice plots above we conclude that stronger motions are required to
increase \bsq at the bottom, as seems very reasonable given the greater density there.
\begin{figure}
\includegraphics[clip,trim=0.5cm 5cm 0.5cm 1cm]{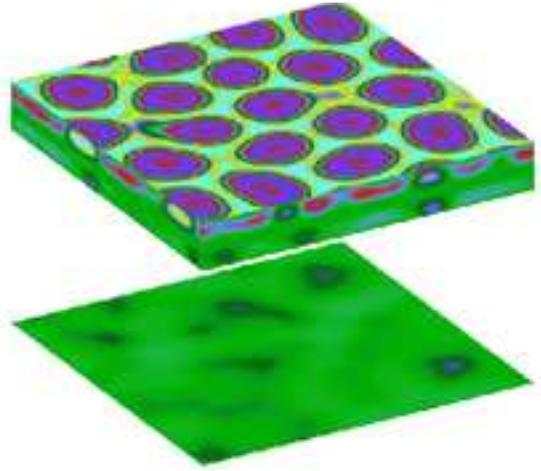}
\caption{Relative distribution of vertical component of magnetic field (near the top and
  bottom) and vertical momentum density (sides), for the case $Q=500$
  at $t=37.12$.}\label{mhd1_3d}
\end{figure}

Having discussed the weak field, $Q=100$ case, we now move to
examine the effect of increasing the Chandrasekhar number to
$Q=500$. As we have shown for the $Q=100$ case, the motion is
largely confined to the top and bottom layers shown in Figure
\ref{mhd1_3d}. Furthermore, for this $Q=500$ case, near the top we
notice hexagonal-type cells of size $\sim 1.6-1.8$ dominate, the
stronger field has reduced horizontal scales because particle motion
is more confined along field lines. The sides of the box also show
that the convection cells in the upper layer are less prominent than
for $Q=100$.
\begin{figure}
\begin{center}
\includegraphics[width=.5\linewidth,clip,trim=0.5cm 6.5cm 0cm 0.75cm]{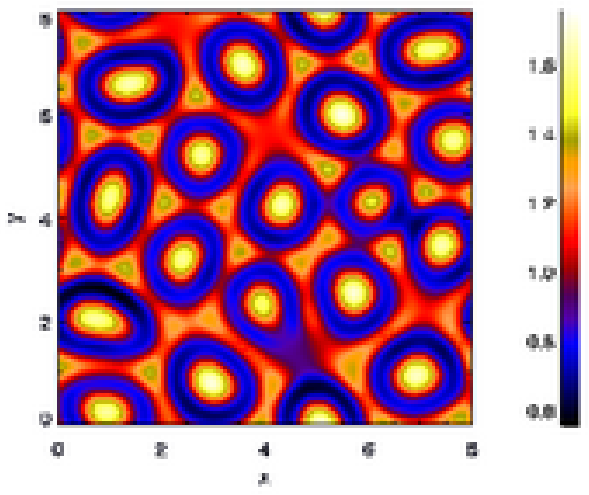}\includegraphics[width=.5\linewidth,clip,trim=0.5cm 6.5cm 0cm 0.75cm]{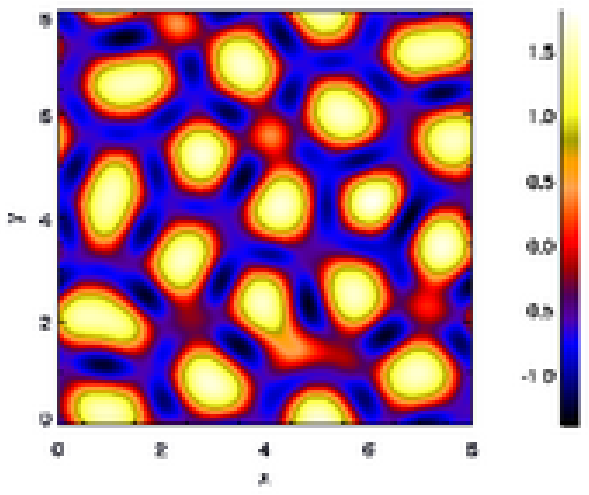}\\
\includegraphics[width=.5\linewidth,clip,trim=0.5cm 6.5cm 0cm 0.75cm]{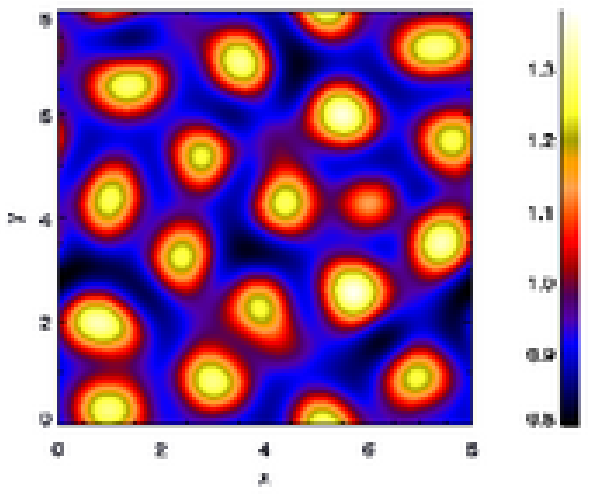}\includegraphics[width=.5\linewidth,clip,trim=0.5cm 6.5cm 0cm 0.75cm]{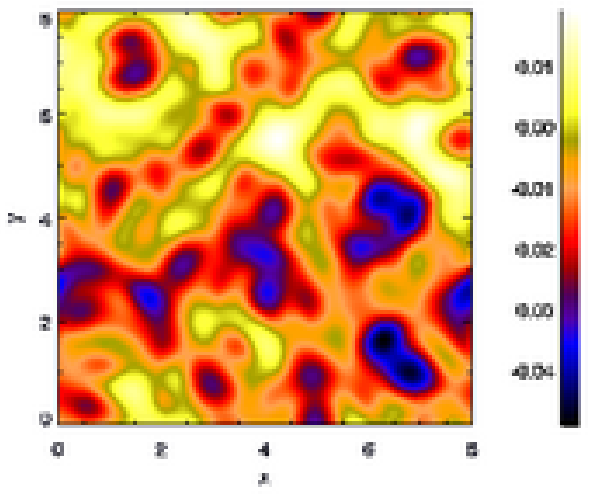}\\
\includegraphics[width=.5\linewidth,clip,trim=0.5cm 6.5cm 0cm 0.75cm]{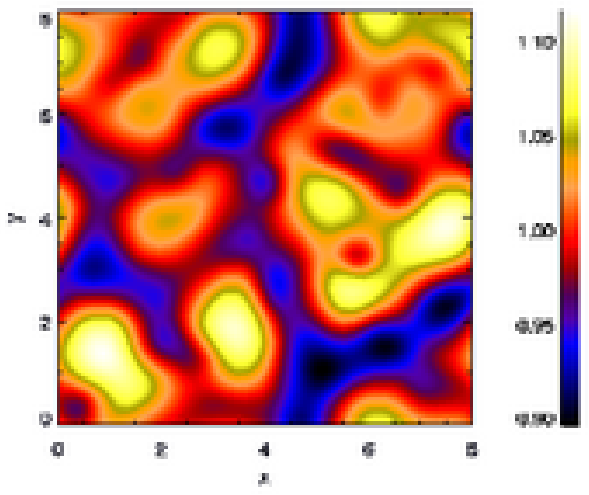}\includegraphics[width=.5\linewidth,clip,trim=0.5cm 6.5cm 0cm 0.75cm]{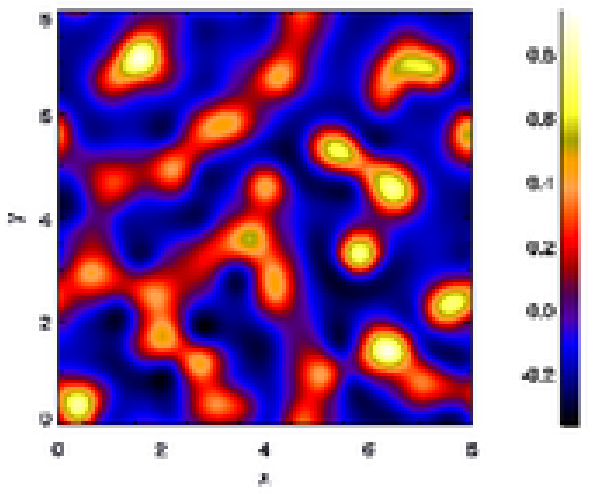}\\
\end{center}
\caption{From top to bottom: vertical component of magnetic field
(left) and vertical component of momentum (right) in the horizontal
plane
  at $z=0.75, 1.5, 2.25$; for $Q=500$ and $t=37.12$. }\label{mhd1_mag_layer}
\end{figure}
The variation with $z$ for the $Q=500$ case is shown more clearly in
Fig. \ref{mhd1_mag_layer}. At $z=0.75$, $B_z$ is concentrated in
circular and triangular cells, corresponding to upflow  and downflow
 regions in the $\rho u_z$ plot. These regions of concentrated
vertical flux are separated by rings  with low $B_z$, and correspond
to regions of low $|\rho u_z|$. At this depth there is strong
correlation between field and motion; a behaviour that is similar to
$Q=100$ and is again consistent with the general picture of
one-layer magnetoconvection. In the stable region, at $z=1.5$, the
associated  disturbance to
 the stronger field has been effectively mirrored or transported well into
the mid-layer. A possible consequence of such magnetic `connection'
between these two layers is that field lines may act as channels and
guide particles (not possible for weak fields) from one region into
another. A typical convection cell would require some horizontal
motion but if this is opposed by a strong vertical field then
particles are more likely to continue in the vertical direction.
However, we must also note that increasing field strength is to
reduce motion, as discussed later, so in terms of overshooting
there is a competition between the two factors.

Fig. \ref{mhd1_mag_layer} shows the middle layer ($z=1.5$) has generally lower values of $B_z$
and around half as much contrast than at $z=0.75$  because motion is
less vigorous and $\rho u_z$ at this depth  is typically an order of
magnitude smaller than at $z=0.75$. The figure shows no correlation
between $\rho u_z$ and $B_z$ for the mid-layer, as with $Q=100$.
However, in contrast to the weak field case where there is some similarity
between $\rho u_z$ at $z=0.75, 1.5$; here the hexagonal structure in
the upper layer is entirely absent in the middle. Despite the strong
similarity in $B_z$, information about vertical motion is not
transported from the top to the middle. In fact, comparing $\rho
u_z$ at $z=1.5,2.25$ show some correspondence (see, for example, the
roll [in red] near the top left of the two plots). This suggests the
increased field may have reduced overshooting from top and increased
it from the bottom. We conclude from Fig. \ref{mhd1_mag_layer} that,
since there is no requirement that $B_z$ and $\rho u_z$ to be
related in a convectively stable region, the middle may echo either of the
structures above or below.
\begin{figure}
\includegraphics[scale=.50,clip,trim=2cm 13cm 0cm 0cm]{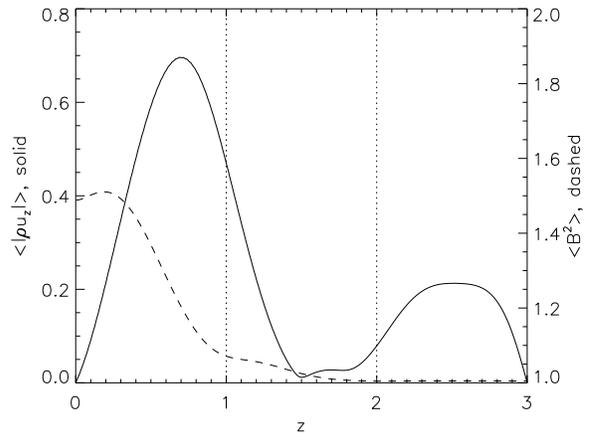}
\caption{Horizontal average of modulus of vertical momentum (solid
line) and
  magnetic energy divided by F (dashed line), as a function of
  depth, for $Q=500$ at $t=37.12$.}\label{mhd1_layer}
\end{figure}
In order to test whether the effect of increasing the Chandrasekhar
number is to reduce overshooting from the top and increase it from
the bottom, we examine once again, the modulus of the vertical component of
momentum density shown in  Fig. \ref{mhd1_layer} to compare the
three regions. In conjunction with Fig. \ref{mhd2_layer} ($Q=100$),
we see that $\langle|\rho u_z|\rangle_{Q=100} > \langle|\rho
u_z|\rangle_{Q=500}$ so increased field has generally suppressed
convection. Although the plot for $Q=500$ is qualitatively very
similar to that for $Q=100$ the decrease in $max(\langle |\rho u_z|\rangle)$
from $Q=100$ to $Q=500$ is $1.18\to 0.70$ in the upper layer and
$0.29\to0.21$ in the bottom layer and thus activity in the top layer is more strongly suppressed.
This implies that the extent of overshooting from the bottom
relative to that from the top has increased with field strength.
Although the top has more vigorous motion, for overshooting we must consider also the 
direction of motion, and we will return to this when we
consider \mom later. Fig. \ref{mhd1_layer} shows that the variation
of \bsq, in comparison to the weak field case, is smaller but that the 
variation with $z$ is similar. The typical value at the middle is
$\sim0.70$ times the maximum, which shows that stronger fields can increase
the amount of magnetic energy pumped downwards, thereby transporting
the structure and hence providing more connection. However, since
the viogour of motion is suppressed (compared to $Q=100$) there is
no increased overshooting from the top.

\begin{figure}
\includegraphics[clip,trim=0.5cm 5cm 0.5cm 1cm]{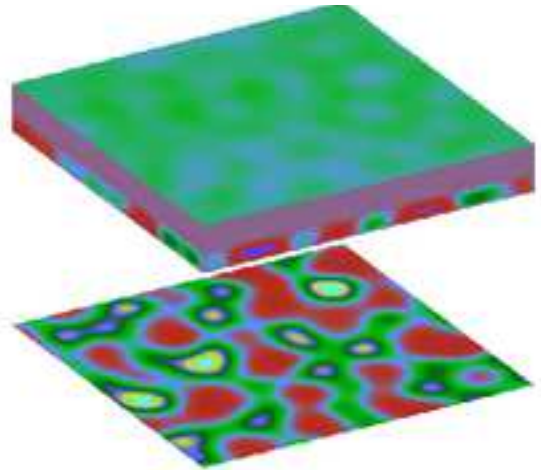}
\caption{Relative distribution of vertical component of magnetic field (near the top and
  bottom) and vertical component of momentum (sides), for the case $Q=1000$
  at $t=53.60$.}\label{mhd3_3d}
\end{figure}
\begin{figure}
\includegraphics[clip,trim=0.5cm 6.5cm 0cm
  0.75cm]{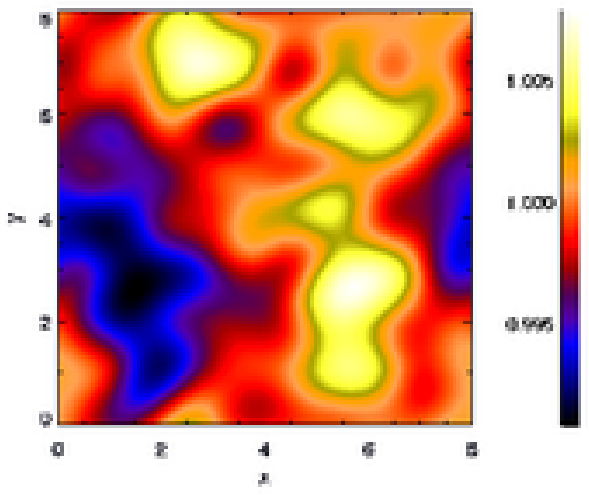}\\
\includegraphics[clip,trim=0.5cm 6.5cm 0cm
  0.75cm]{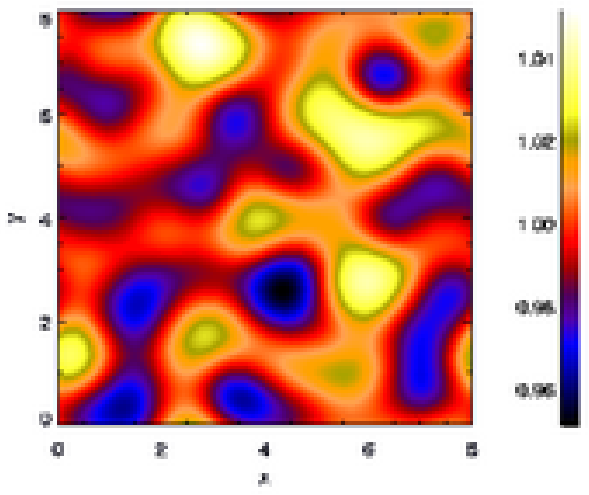}\\
\includegraphics[clip,trim=0.5cm 6.5cm 0cm
  0.75cm]{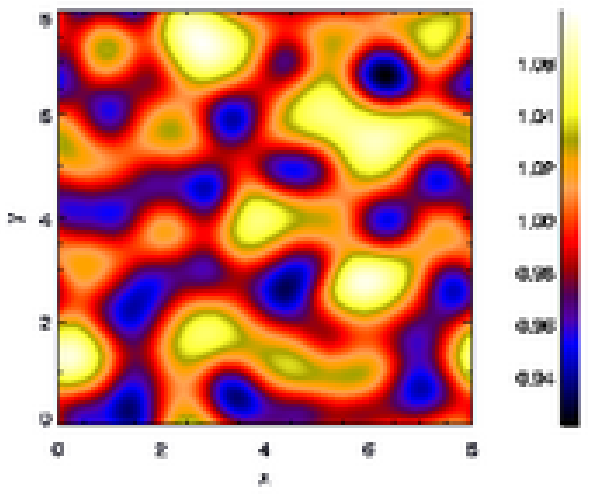}\\
\caption{From top to bottom: vertical component of magnetic field in the horizontal plane
  at $z=0.75, 1.5, 2.25$; for $Q=1000$ and $t=53.60$. }\label{mhd3_mag_layer}
\end{figure}
\begin{figure}
\includegraphics[scale=.50,clip,trim=2cm 13cm 0cm 0cm]{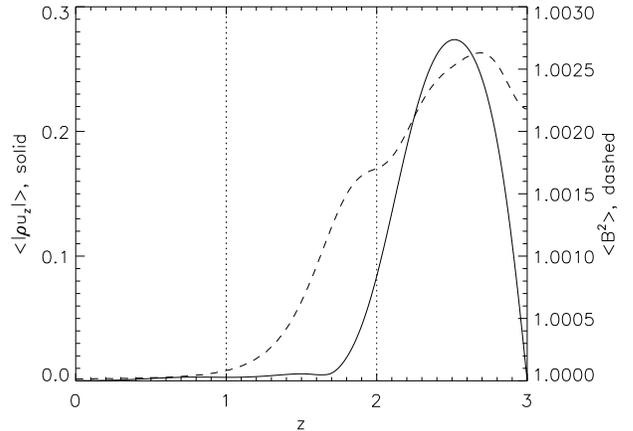}
\caption{Horizontal average of modulus of vertical momentum (solid)
and
  magnetic energy divided by F (dashed line), as a function of
  depth, for $Q=1000$ at $t=53.60$.}\label{mhd3_layer}
\end{figure}

We now move to discuss the $Q=1000$ case for which Fig. \ref{mhd3_3d} shows an almost inverted 
distribution of structure and activity compared with that for $Q=100$ and $Q=500$.
The bottom has
 magnetic structure with a horizontal scale comparable to that for the top of the $Q=500$ case, although the distribution is less ordered. Convection
is predominantly in the bottom layer with typical size $\sim2.2$
units. The strong applied field has caused the top to be almost
featureless with an almost uniform $B_z$ and little vertical motion.
The slice plots shown in Fig. \ref{mhd3_mag_layer} confirm this
effect. This figure shows that the form of $B_z$ at $z=0.75$ is different to that at $z=1.5,
2.25$ but the latter two plots are similar; the disturbance to $B_z$
is now transported from the bottom upwards, in contrast to the case $Q=500$. Although the
plots show rich structure,  note that the contrast in $B_z$ is only
$\sim 0.01, 0.05, 0.12$ units for $z=0.75,1.5,2.25$ respectively, and
these are all smaller than for previous cases, so the field is almost
unperturbed and remains mostly vertical. However, as for previous
cases we found the strongest flow-field correlation for the layer
with most magnetic disturbance, which is the bottom layer for
$Q=1000$. This continues the trend from $Q=100\to500$, that
overshooting from the bottom relative to the top  has increased, and
is further supported by Fig. \ref{mhd3_layer} that show vertical
motion almost completely suppressed for $z<1.7$ but $\langle|\rho
u_z|\rangle$ is still comparable to previous cases in the lower
convection zone. Since the top is suppressed, overshooting from the
bottom dominates; in fact $\langle|\rho u_z|\rangle$ for $z>2$ is
qualitatively similar to the reverse of the curve in the top layer
in $Q=100$ and $Q=500$.  A strong field resists deformation so there
is only a 1\% perturbation to \bsq. As before, the most vigorous region
(bottom layer here) contains most magnetic energy but unlike
previous cases the middle has a significant portion of magnetic
energy. These observations are again different from that for
$Q=100,500$. We have also done calculations for $Q=1500$; these show the same effect as for $Q=1000$ but to a greater degree, and for $Q=750$, which show features intermediate between the $Q=500$ and $Q=1000$ cases.

\section{Conclusions}\label{conc}

In this letter we have examined three-layer magnetoconvection 
and we have focused on the effect of varying the
strength of the magnetic field via varying the Chandrasekhar number
$Q$.  For weak imposed magnetic field and for our parameter choices convection occurs in both
the top and bottom layers. For such fields the magnetic field behaves
passively and is easily swept into the intracellular regions. As we
increased the strength of the magnetic field we showed at the magnetic
field forces substantial changes onto the flow. We showed that for
modest strength magnetic field, e.g. for the $Q=500$ case, the
magnetic field forces a fairly regular convection pattern in the upper
layer. However, we showed that if the magnetic field
becomes too strong, as for example in the $Q=1000$ case motion in the
upper convection zone is almost completely suppressed.

This preliminary work has provided the first steps towards
understanding the effects of imposed magnetic fields on a stellar
atmosphere with multiple unstable regions. Although the geometry is
somewhat idealised, the results do show that the efficiency of
overshooting and the way in which the unstable layers can communicate
with each other through a stable region can be significantly affected
by a magnetic field permeating all three layers. Further
work at higher Rayleigh numbers is undoubtedly required to show
whether the behaviour found persists in turbulent flows.  

\ack

LJS \& MREP wishes to thank STFC for the award of a rolling grant to fund research in
magnetoconvection. MKL wishes to thank St.\ Catharine's College for support
for this project. We are grateful to Paul Bushby for helpful discussions.

\end{document}